# Strain-tunable entangled-light-emitting diodes with high yield and fast operation speed

*Jiaxiang Zhang [1] \*, Johannes S. Wildmann [2], Fei Ding [1] \*, Rinaldo Trotta [2] \*, Yongheng Huo [1], Eugenio Zallo [1], Daniel Huber [2], Armando Rastelli [2] & Oliver G. Schmidt [1]*

[1] Institute for Integrative Nanosciences, IFW Dresden, Helmholtzstraße 20, 01069 Germany

[2] Institute of Semiconductor and Solid State Physics, Johannes Kepler University Linz,

Altenbergerstraße 69, 4040, Austria

**Triggered sources of entangled photons play crucial roles in almost any existing protocol of quantum information science [1, 2, 3, 4]. The possibility to generate these non-classical states of light with high speed and using electrical pulses could revolutionize the field. Entangled-light-emitting-diodes (ELEDs) based on semiconductor quantum dots (QDs) are at present the only devices that can address this task [5]. However, ELEDs are plagued by a source of randomness that hampers their practical exploitation in the foreseen applications: the very low probability (~$10^{-2}$) of finding QDs with sufficiently small fine-structure-splitting for entangled-photon-generation [6]. Here, we overcome this hurdle by introducing the first strain-tunable ELEDs (S-ELEDs) that exploit piezoelectric-induced strains to tune QDs for entangled-photon-generation. We demonstrate that up to 30% of the QDs in S-ELEDs emit polarization-entangled photon pairs with entanglement-fidelities as high as $f^+=$**

**0.83(5). Driven at the highest operation speed of 400 MHz ever reported so far, S-ELEDs emerge as unique devices for high-data rate entangled-photon applications.**

Analogous to the cascade emission in atomic systems [7], electrical injection of electrons and holes into an ELED containing QDs switches on the radiative decay of biexciton (XX) to the exciton (X) to the crystal ground state (0) [8]. The two ideally-degenerate intermediate X states (spin ±1) lead to two indistinguishable decay paths with the ensuing emission of polarization-entangled photon pairs. The two-photon quantum-mechanical state can be expressed with the Bell state $|\Psi^+> = 1/\sqrt{2}\ (|H_{XX}H_X> + |V_{XX}V_X>)$ (H and V denote the orthogonally horizontal and vertical polarizations). In real QDs, however, a reduced structural symmetry due to the anisotropy in strain, composition and shape results in the appearance of an energetic splitting between the two bright X states, the so-called fine structure splitting (FSS, $s$) [9, 10]. In the presence of the FSS, the entangled state evolves over the X lifetime and the time-averaged fidelity to the Bell state reveals classical correlations among the emitted photons. High-fidelity to the state $|\Psi^+>$ can be observed only by temporal post-selection of the emitted photons that, however, results in a strong reduction of the brightness of the quantum source. Therefore, the FSS in semiconductor QDs is the key parameter determining the quality of the entangled photon pairs. Recent work shows that in standard self-assembled QDs [6], the probability of finding QDs with a FSS smaller than the radiative linewidth of the X emission (1 µeV) is $< 10^{-2}$. This finding implies that as-grown QDs are still unpractical for scalable quantum networks. For example it has been found that only 1 QD per ELED shows a FSS small enough for entangled-photon generation [5]. In this context, the real potential of ELEDs for entangled-photon generation can be harnessed only when a tight control over the FSS is achieved.

The FSS can be suppressed or tuned to zero via the application of either a vertical electric field [11] or a combination of strain and electric field [12, 13]. The main drawback of these approaches is the difficulty of using the electric field to control the FSS and to inject carriers simultaneously, as in ELEDs. Other techniques such as thermal annealing [14, 15] and in-plane magnetic field [16] would be potentially compatible with the ELEDs, but the former requires a lengthy procedure and the latter bulky setups, thus rendering a practical implementation inconvenient. Recent theoretical works (although not experimentally realized so far) suggest that the FSS can be eliminated using solely a well-controlled strain field [17, 18]. Using piezoelectric-induced strains to engineer the properties of ELEDs would be highly desirable because this fully electro-mechanically-controlled tuning knob would allow the problems related to the FSS in ELEDs to be overcome.

Here we experimentally demonstrate such a quantum device by integrating ELEDs onto a piezoelectric actuator featuring giant piezo-electric response and capable of delivering well-defined anisotropic strain fields. With this novel device – which we call S-ELEDs – we show (*i*) that the FSS of QDs can be tuned effectively with the elastic strain fields without affecting the electrical injection of the operation of the ELEDs; (*ii*) that up to 30% of the QDs are tuned to be suitable for the generation of entangled-photon pairs (more than an order of magnitude more than in previous devices [5]) and (iii) the highest operation speed ever reported so far for an entangled-photon source (400 MHz). This unique set of properties paves the way towards the real exploitation of ELEDs in high data-rate entangled-photon applications involving a large numbers of quantum emitters.

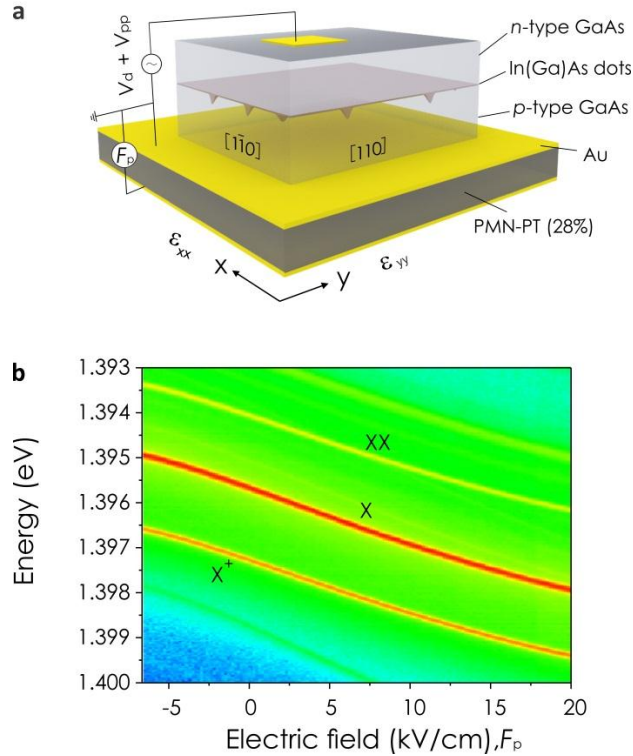

**Figure 1 | Strain-tunable entangled-light-emitting diode. a**, Sketch of the diode structure. Different from previous works, the PMN-PT top surface has (011) orientation, which imposes large anisotropic strain fields with well-defined orientation onto the overlying ELED. **b**, Electroluminescence from a single QD in an S-ELED versus electric field $F_p$ applied to the PMN-PT actuator.

The S-ELED studied in this work is schematically shown in Fig. 1a. A 440 nm-thick *n-i-p* nanomembrane containing InGaAs QDs is integrated onto a 0.3 mm-thick [Pb(Mg$_{1/3}$Nb$_{2/3}$)O$_3$]$_{0.72}$[PbTiO$_3$]$_{0.28}$ (PMN-PT) single piezoelectric crystal. The detailed fabrication process is described in *Methods*. Different from previous works dealing with strain tuning of QDs via PMN-PTs [12, 13, 19, 20], the actuator used here has pseudo-cubic cut directions [100], [0-11] and [011], denoted by *x*, *y*, *z* axis, respectively. When the PMN-PT is poled along the *z* axis, in-plane strain fields with normal components $\varepsilon_{xx}$ along the *x* axis and $\varepsilon_{yy}$ along the *y* axis with opposite sign can be transferred to the nanomembrane. Accounting for its relevant piezoelectric coefficients $d_{31}$ ~ +420 pC/N along the *x* axis and $d_{32}$ ~ -1140 pC/N along the *y* axis [21], the in-plane anisotropy is estimated to be $\varepsilon_{xx} \approx -0.37\varepsilon_{yy}$. The large and well controlled strain anisotropy and the broad range of attainable strain magnitudes are unique and turned out to be vital in our work.

Apart from the strain fields, electrical contacts are arranged in such a way that electrical fields can be independently applied across the diode and the PMN-PT actuator. By biasing the diode and applying a variable electric field ($F_p$) to the PMN-PT simultaneously, energy-tunable electroluminescence (EL) from a single QD is produced, as shown in Fig. 1b. According to the power and polarization-resolved measurements the observed EL lines are ascribed to exciton, biexciton and charged exciton emission, respectively. As the magnitude of $F_p$ is varied, all the emission lines shift in energy. The shift is expected to be mostly proportional to the opposite of the variation of the volumetric strain at the QD position, which can be estimated as $\varepsilon_{tot} = (1 - 0.37)\varepsilon_{yy}\frac{S_{11}+2S_{12}}{S_{11}+S_{12}} \approx 0.33\varepsilon_{yy} \approx 0.33d_{32}F_p$ (with $S_{ij}$ the compliance coefficients of the host material). Since $\varepsilon_{tot}$ has the same sign as $d_{32}$, which has a relatively large magnitude and is negative, we expect a positive $F_p$ to induce a compressive strain which results in a blue shift of the EL, while a negative $F_p$ induces a tensile strain, which results in a red shift. We also note that a total energy shift of about 2.5 meV is achieved as $F_p$ is varied from -6.7 to 20 kV·cm$^{-1}$.

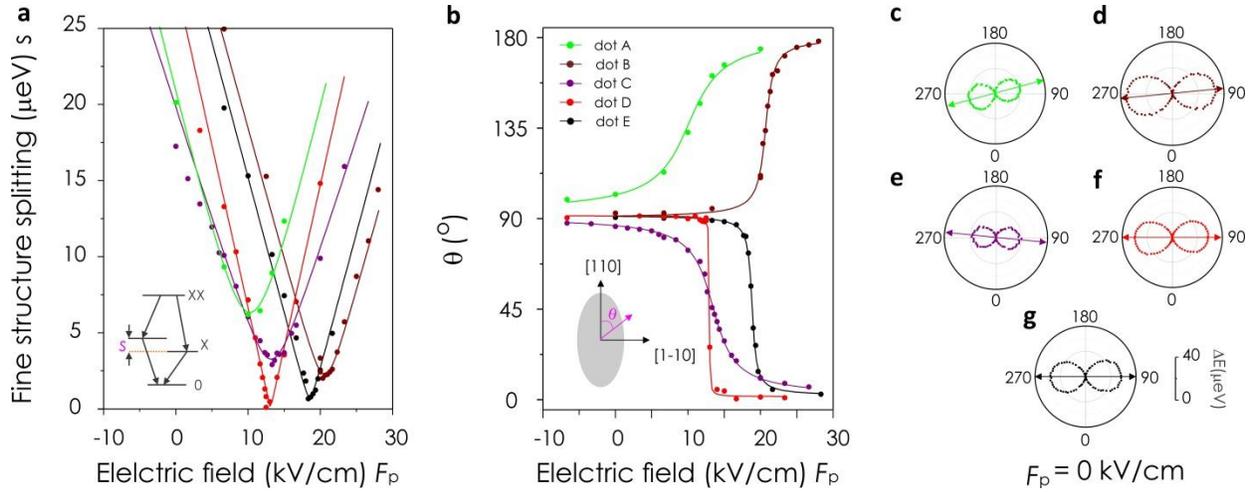

**Figure 2 | Strain-induced change of fine structure splitting and exciton polarization angle.**
**a**, **b**, Representative variation of the FSS (*s*) and the polarization direction $\theta$ of the high-energy component of the exciton as a function of $F_p$ for five QDs. The insets show sketches of biexciton cascade and the orientation of the exciton polarization. **c**, **d**, **e**, **f**, **g**, $s_0$ and $\theta_0$ corresponding to the studied five QDs at $F_p = 0$ kV·cm$^{-1}$. In the polar plot $0^0$ corresponds to the [110] and $90^0$ to the [1-10] crystal axis of the GaAs nanomembrane.

In order to control the FSS of the QDs embedded in the diode, the crystal axes [1-10] and [110] of the GaAs nanomembrane were carefully aligned along the *x* and *y* axes of the PMN-PT actuator, respectively (see Fig. 1a). Representative plots of the FSS of different QDs as a function of $F_p$ are shown in Fig. 2a. Although the emission energy shift is only about 2.5 meV as $F_p$ is varied from - 6.7 to 28 kVcm$^{-1}$ as a consequence of the strong strain anisotropy (and thus relatively small $\varepsilon_{tot}$), the FSS is tuned over a broad range from 30 ~ 0 μeV. Away from the minimum FSS, all studied QDs exhibit an approximately linear change in the FSS with $F_p$ at a rate of about 2.0 μeV·kV$^{-1}$·cm, which is about 9 times larger than what has been reported for vertical electric fields [11]. In addition to this drastic change in FSS, we also observe rotations of the exciton polarization angle $\theta$, i.e., the polarization direction of the high-energy line of the exciton with respect to the [110] direction of the GaAs nanomembrane (see Fig. 2b). At the largest available tensile (compressive) strain, $\theta$ for all QDs tends to be directed along the [1-10] ([110]) direction. Furthermore, we note that the above tuning behavior is mainly determined by the exciton polarization angle at zero strain fields ($\theta_0$) with respect to the predefined direction of the strain [12, 22]. Experimentally this polarization angle can be extracted from $\Delta E = |E(\theta, F_p = 0) - E_{min}|$ as a function of $\theta$, where $E_{min}$ is the minimum energy of $E(\theta, F_p = 0)$. Fig. 2c ~ g show the dependence of $\Delta E$ on $\theta$ for the five QDs. The initial FSS $s_0$ and $\theta_0$ are represented by the magnitude and orientation of the lobes in each polar plot. For the dot A and B, $\theta_0$ are 102.0° ± 0.4° and 92.7° ± 0.2°, respectively, suggesting positive deviations of $\theta_0$ from the strain $x$ axis. Consequently, as $F_p$ is increased, the polarization angle $\theta$ rotates counterclockwise and the FSS experiences a finite lower bound $s_{min}$= 7.2±0.2 µeV and 2.2±0.1µeV (the green and wine curves in Fig. 2a and 2b). For dot C, $\theta_0$ is found to be 86.1° ± 0.4°, which suggests a negative deviation from the strain $x$ axis. Thus a clockwise rotation of $\theta$ over $F_p$ is observed, together with a $s_{min}$= 4.2±0.2 µeV (the purple curve). For the QDs D and E, $\theta_0$ is found to be 90.4° ± 0.3° and 90.6° ± 0.3° respectively, which indicate exact alignment of $\theta_0$ and the strain $x$ axis. The polarization angles for both QDs rotate clockwise as the strain is varied, and this implies that their polarization angles at zero strain fields are oriented at angles slightly less than 90°. It should be noted that this difference from the observed values (larger than 90°) is ascribed to the limited alignment precision of the polarizer (within a few degrees) [12]. Most importantly, owing to this exact alignment between the exciton polarization angle $\theta_0$ and the strain axes for dot D and E, their FSS can be reduced well below 1 µeV and $s_{min}$ is found to be 0.30 ±0.25µeV and 0.60± 0.20 μeV respectively. By simply treating the anisotropic strain as an effective uniaxial strain [18], the behavior of $s$ and $\theta$ are well fitted (solid lines in Fig. 2a and 2b). Noticeably, the minimum reachable FSS value, following the form $s_{min} = s_0|\sin(2\theta_0)|$, is determined only by $s_0$ and $\theta_0$ (see *Supplementary Information*). The $s_{min}$ predicted before varying the strain fields are: 8.18±0.20 μeV, 2.34±0.21 μeV, 3.3±0.18 μeV, 0.24±0.32 μeV and 0.76±0.15 μeV for dot A, B, C, D, E, which show excellent agreement with our experimental data. From our experimental observations and theoretical analysis, it is clear that, in order to cancel the FSS with our external strain fields, the polarization angle at zero applied stress should be as close as possible to the strain principal axes.

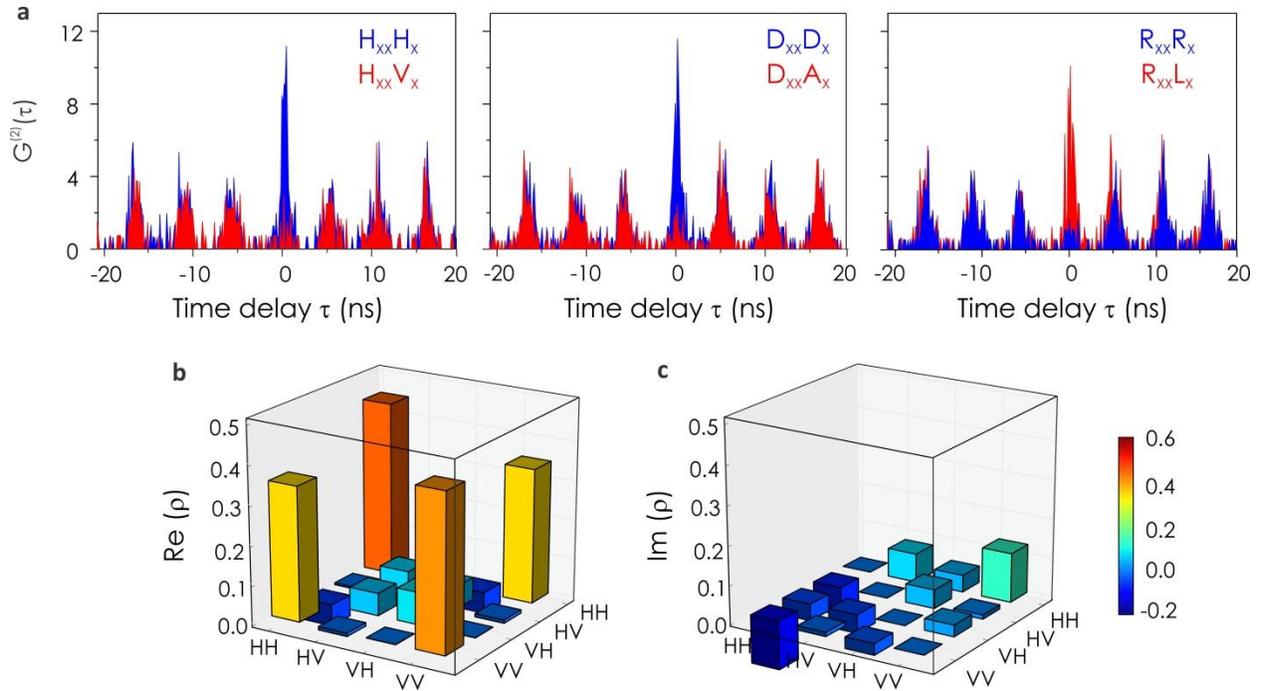

**Figure 3 | S-ELED as source of polarization entangled photons. a**, co-polarized (blue) correlation and cross-polarized (red) correlation counts for a QD in an S-ELED excited with electrical pulses with 185.2 MHz repetition rate, measured in the rectilinear, diagonal and

circular bases. Representative density matrix: **b**, real part and **c**, imaginary part, which is reconstructed with 16 coincidence counts integrated in a 1.8 ns temporal window centered at zero delay time.

The ability to tune the FSS of the QDs to zero allows us to investigate the capability of the S-ELED to generate polarization-entangled photon pairs without the aid of post filtering techniques [23, 24]. Fig. 3a shows the polarization resolved co- and cross-polarization correlation between the XX and X photons emitted by dot E under electrically pulsed excitation and tuned to a FSS of $0.60 \pm 0.20$ μeV (at $F_p = 18.7$ kV·cm$^{-1}$). The periodic correlation peaks with well-separated temporal distance of 5.4 ns arise from the chosen repetition rate of 185.2 MHz. Most importantly, for co-polarized two-photon, strong correlations are observed in linear (HV) and diagonal (DA) bases, while strong anti-correlations are observed in circular basis (RL), as expected for the photon pairs emitted in the maximally entangled Bell state $|\Psi^+\rangle$. In order to quantify the degree of polarization entanglement, we have reconstructed the two-photon density matrix by performing quantum state tomography measurements, as described in ref. 25. Sixteen polarization correlation measurements were performed and the density matrix is reconstructed using the Maximum Likelihood Estimation. The imaginary part and the real part of the density matrix are displayed in Fig. 3b and 3c. The outer off-diagonal elements in the real part of the density matrix reveal a high probability for a superposition of the two-photon wave function, being a clear signature of polarization entanglement [26]. Specifically, the density matrix can be used to quantify the degree of entanglement by extracting the tangle T, the concurrence C, the largest eigenvalue $\lambda$ and the Peres criterion P. We find C = 0.688 ± 0.040 (> 0), T = 0.474 ± 0.055 (> 0), $\lambda$ = 0.795 (> 0.5) and P = -0.30 ± 0.02 (< 0). All these tests exceed the classical limit, proving that the quantum state obtained in our experiment is highly entangled. Using the largest eigenvalue, we are able to determine the most probable state of the system: $1/\sqrt{2}\,(|H_{XX}H_X>$ $+e^{-0.11\pi i}|V_{XX}V_X>)$, in which the presence of the phase $\phi_0 = -0.11\pi$ is likely due to the reflection at the beam splitter [13]. As a consequence, the fidelity to the maximally entangled Bell state $|\Psi^+>$ is found to be $f^+ = 0.766 \pm 0.051$.

Having demonstrated generation of entangled-photon pairs from our S-ELED, we present one of the most important results of our work, that is, the capability of anisotropic strain fields to tune about 30% of the QDs for entangled-photon emission. A statistical study from 82 randomly selected QDs revealed that the majority of QDs in our S-LED device have $\theta_0$ oriented close to the [1-10] crystal axis (see *Supplementary Information*). Therefore, the alignment of the strain axes parallel (or perpendicular) to the [1-10] crystal axis of the GaAs ensures frequent observation of QDs with low value of $s_{min}$. Fig. 4a shows the statistical investigation of $s_{min}$, falling into a range of 0 ~ 40 µeV. Remarkably, 9 QDs revealed a $s_{min}$ less than 1 µeV, which is of the order of the homogenous linewidth of the X emission. Therefore 11% of dots have sufficiently small FSS for strong entanglement in our device. In addition, recent works have suggested that for InGaAs QDs entanglement is robust and the violation of the classical limit can be achieved for FSS smaller than 3 - 4 µeV [13, 27]. In order to quantify this probability in our S-ELEDs, we use the following approach: we study the evolution of the fidelity to the maximally entangled Bell state $|\Psi^+>$ as a function of the FSS for one single QD, and we use the obtained result to estimate at which value of the FSS it is possible to overcome the classical limit. To determine the entanglement fidelity, polarization correlations were performed for each value of the FSS and entanglement was equivalently quantified by measuring the degree of correlation (see *Methods*). As shown in Fig. 4b, the maximum fidelity $f^+ = 0.75 \pm 0.02$ is achieved when the FSS is tuned close to zero. For FSS values larger than 3 µeV, the fidelity drastically decreases below the classical limit (see dashed line). Taking into account that the exciton lifetime of the InGaAs QDs in our S-ELED device has typical values of about 1 ns, this FSS of about 3 μeV provides an upper limit to observe entanglement, consistent with previous reports for InGaAs QDs [13, 27, 28, 29]. From the statistical investigation we find that 27 QDs can be tuned below 3 μeV, which indicates a probability as high as 33% of QDs that can be exploited as entangled-light emitters in our S-ELEDs. Compared to the only work on ELEDs present in the literature [5], this is more than an order of magnitude (a factor ~30) higher. This probability is even higher than what was reported for highly symmetric pyramidal QDs where, however, electrical injection has not been realized yet [30].

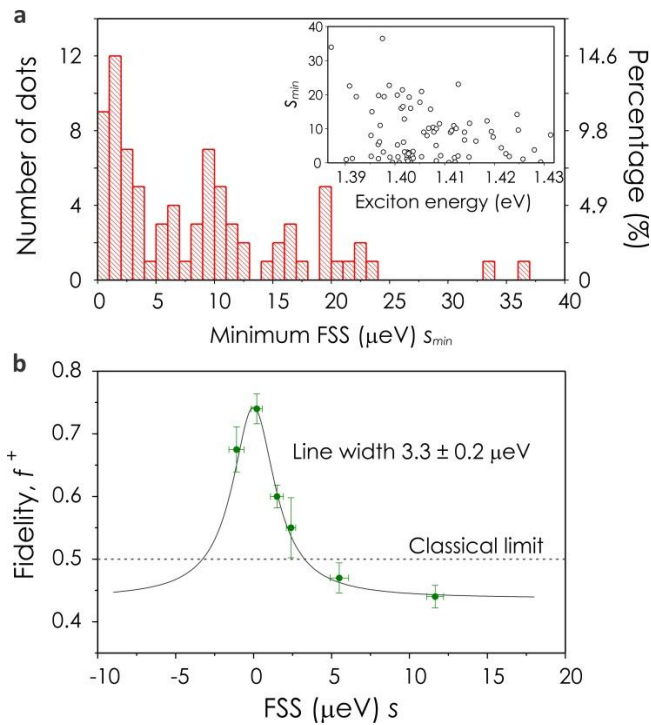

**Figure 4 | Statistical investigation of the minimum FSS and dependence of the entanglement on the value of FSS. a**, Histogram of the distribution of $s_{min}$ tuned by the externally induced strain fields in the S-ELED device. The inset is a scatter plot of $s_{min}$ as a

function of X emission energy. $s_{min}$ shows no energy dependence. **b**, Fidelity as a function of FSS dynamically tuned by the anisotropic strain fields and the solid line is Lorentzian fit with a full width at half maximum of 3.3 ± 0.2 µeV.

In addition, we can increase the pulsed excitation rate in order to enhance the generation rate of the entangled-photon pairs. This feature is highly desirable for high data-rate quantum information processing. Fig. 5 shows the results of polarization correlation measurements at 400 MHz for the dot E. Similar to the case of 185.2 MHz, we observe correlations in the HV and DA bases and anti-correlation in the RL basis for co-polarized two photons. In the left panel of Fig. 5b the degrees of correlation in given bases are reported. We find a state fidelity as high as 0.66 ±0.02, which exceeds the classical limit of 0.5 and thus proves, for the first time, generation of entangled-photon pairs at 400 MHz. We observe that the fidelity at 400 MHz is smaller than the fidelity at 185.2 MHz reported above. This is likely ascribed to the contribution of a small amount of uncorrelated photon pairs due to the time-dependent re-excitation process, residual FSS, and background emission [5, 26, 27]. By temporal post-selection of the emitted photons we can alleviate these effects for the entanglement degradation. The red curves in Fig. 5b show the degrees of correlation for a temporal gate Δτ = 0.8 ns at which ~20% of the coincidence counts are discarded. We measure $C_{HV}$ = 0.67±0.06, $C_{DA}$ = 0.63±0.04 and $C_{RL}$ = -0.78±0.07, corresponding to $f^+$ = 0.77±0.03. The degree of correlation and entanglement fidelity can be improved by further shortening Δτ [5, 26, 27, 29, 31, 32]. With the narrowest available gate width of 0.1 ns applied, ~80% coincidence counts are discarded and the degrees of correlation increase significantly: $C_{HV}$ = 0.74±0.12, $C_{DA}$ = 0.74±0.09 and $C_{RL}$ = -0.84±0.12, which provides the highest fidelity of 0.83±0.05. It is interesting to investigate whether such a high level of entanglement is sufficient to violate Bell's inequality. Using the measured values of the degree of polarization correlation, it is possible to determine Bell parameters: $S_{RD}$, $S_{RC}$ and $S_{DC}$, which are related to three different planes of the Poincaré sphere (see *Methods*). Our results show that the Bell parameters increase as the gate width is decreased, and the Bell inequality is found to be violated starting from the gate width of 0.8 ns. From the degrees of correlation at $\Delta\tau = 0.8$ ns, we calculate $S_{RD} = 1.83 \pm 0.07$, $S_{RC} = 2.04 \pm 0.09$ and $S_{DC} = 2.00 \pm 0.08$. Two of these values are larger than 2 and they indicate violations of the Bell inequality. In particular, $S_{RD}$ is found to be less than $S_{RC}$, $S_{DC}$ and this is due to the weaker degree of correlation observed in the HV and DA bases ($C_{HV}$, $C_{DA} < |C_{RL}|$) [5, 13, 31]. This common feature is usually observed for the QDs entangled-photon sources and is most probably ascribed to the weak coupling between the two bright exciton states [31]. Therefore, for the actual entangled state $S_{RD}$ is not optimally chosen to inspect violation of the Bell inequality. Additionally, for the gate width of 0.1 ns, we find $S_{RD} = 2.09 \pm 0.21$, $S_{RC} = 2.23 \pm 0.21$ and $S_{DC} = 2.23 \pm 0.24$ (see *Supplementary Information*). All these three parameters are above the threshold of 2, thus proving that our S-ELED is capable of generating non-local states of light in response to an electrical trigger.

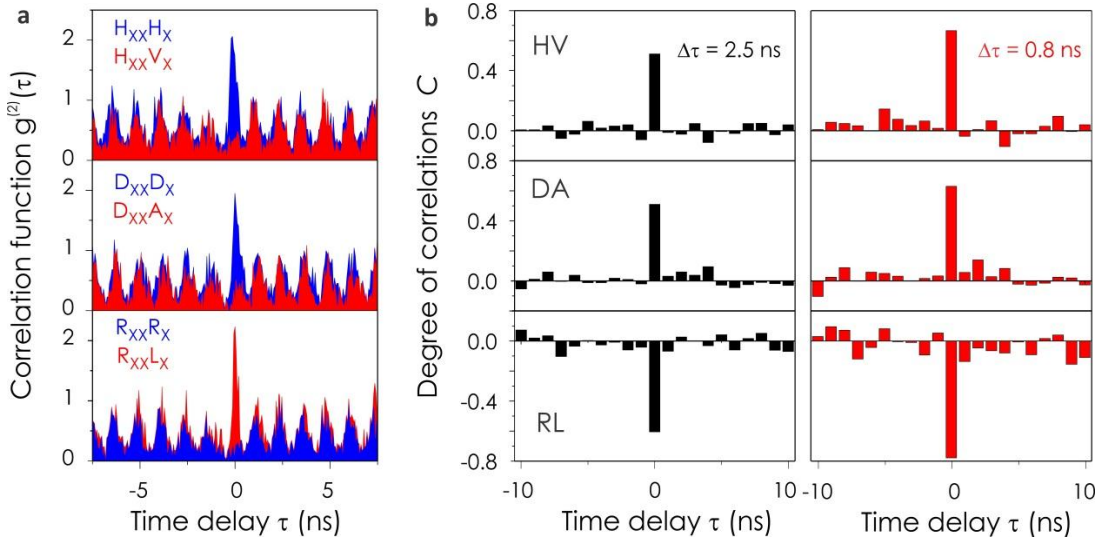

**Figure 5 | Polarization correlation results from the S-ELED under electrically pulsed injection at repetition rate of 400 MHz. a**, Normalized correlation functions for co- and cross-polarized XX and X photons in HV, DA and RL bases. **b,** Degree of correlation $C$ in given basis, in which correlation in HV and DA bases ($C > 0$) and anti-correlation ($C < 0$) in RL basis are obtained without temporal gate ($\Delta \tau = 2.5$ ns) and with a temporal gate width 0.8 ns centered at zero delay time.

In conclusion, we have presented the first S-ELEDs in which anisotropic strain fields are used to tune QDs for entangled-photon generation. We have shown that up to 30% of QDs embedded in this device are capable of emitting polarization entangled-photon pairs. This practically removes the tedious search for special QDs plaguing previous ELEDs [5]. Furthermore, we prove for the first time triggered entangled-photon emission at the repetition rate of 400 MHz, the highest value ever reported so far. Our all electrically-controlled S-ELEDs emerge as the most practical and fastest source of entangled photons available to date, that can be exploited for large-scale quantum communication and quantum computation tasks.

**Methods**

The studied sample was grown on a (001) GaAs substrate by solid-source molecular beam epitaxy (MBE). It consisted of a *p-i-n* heterostructure diode composed of a 178 nm-thick *n*-type GaAs layer, a 160 nm-thick intrinsic GaAs layer and a 96 nm-thick *p*-type GaAs layer from the bottom to the top. A layer of low density ($10^6 \sim 10^7$ cm$^{-2}$) self-assembled InGaAs QDs was embedded in the middle of the intrinsic GaAs layer. The entire diode structure was grown on a 100 nm-thick Al$_{0.75}$Ga$_{0.25}$As sacrificial layer. As for the device processing, first of all, standard

UV photolithography and wet chemical etching were used to fabricate mesa structures with size of $120\times160$ μm$^2$. The longer edge of the GaAs membrane was processed along [110] crystal axis of GaAs and – during the transfer onto the piezoelectric actuator – was carefully aligned along the *y* axis of the PMN-PT actuator. It is worth noting that the bonded gold layer on the bottom formed a *p*-contact, while the *n*-type contact was formed by depositing a gold pad with size of $50\times50$ μm$^2$ on the top of the nanomembrane.

The EL is observed when the diode is biased with a DC voltage ($V_d$) above -1.7 V; however, it is slightly different from one device to another due to the different bonding conditions. The pulsed electrical excitation is accomplished by superimposing an ultrafast electrical pulse stream onto a -1.6 V DC bias by using a broad bandwidth bias-Tee. The pulse stream used in this work has nominal duration of 300 *ps*, and amplitude of $V_{pp}$ = -8.0 V. For optical measurements, the EL emitted from the diode is collected by a 50× microscope objective with numerical aperture of 0.42, which is placed on the top of the nanomembrane and collects the photon emission from the area close to the metal contact. By inserting a half-wave plate and a linear polarizer directly after the collection lens, Polarization-resolved measurements were performed to obtain the FSS vs. $F_p$. The exciton polarization is determined by aligning the fast optical axis of the polarizer along [110] direction of the nanomembrane. The EL was directed to a spectrometer with 750 mm focus length, and the spectrum is analyzed using a nitrogen cooled charge-coupled device. The FSS is determined with an accuracy of sub-μeV by taking the experimental procedure in ref. 11 and 12.

Regarding the polarization resolved correlation measurements, a non-polarizing 50:50 beam splitter is placed directly after the collection objective in order to divide the optical paths between two spectrometers, which are used to detect X and XX separately. After each spectrometer, a Hanbury-Brown Twiss setup, consisting of a polarizing beam splitter and two high efficiency single-photon avalanche detectors, is placed. Half- and quarter-wave were used to select the proper polarization basis. The temporal resolution of the system is about 400 ps. The entanglement can be equivalently quantified by measuring degree of correlation C, which is defined by

$$C_{basis} = (g^{(2)}_{XX,X}(\tau) - g^{(2)}_{XX,\bar{X}}(\tau))/(g^{(2)}_{XX,X}(\tau) + g^{(2)}_{XX,\bar{X}}(\tau)),$$

where $g^{(2)}_{XX,X}(\tau)$ and $g^{(2)}_{XX,\bar{X}}(\tau)$ are normalized second-order time correlations for co-polarized and cross-polarized XX and X photons, respectively. The fidelity $f^+$ is calculated by using the formula: $f^+ = (1 + C_{HV} + C_{DA} - C_{RL})/4$, in which $C_{HV}$, $C_{DA}$ and $C_{RL}$ are degree of correlations in HV, DA and RL bases. The Bell parameters are determined with the formulas: $S_{RD} = \sqrt{2}(C_{HV} + C_{DA})$; $S_{RC} = \sqrt{2}(C_{HV} - C_{RL})$; $S_{DC} = \sqrt{2}(C_{DA} - C_{RL})$.

**References**

1. Bouwmeester, D., Pan, J.-W., Mattle, K., Eibl, M., Weinfurter H. & Zeilinger, A. Experimental quantum teleportation, *Nature* **390,** 575-579 (1997)
2. Knill, E., Laflamm R. & Milburn, G. J. A scheme for efficient quantum computation with linear optics, *Nature* **409**, 46-52 (2001)
3. Kok, P. et al. Linear optical quantum computing with photonic qubits. *Rev. Mod. Phys.* **79,** 135–174 (2007).

4. Gisin, N. & Thew, R. Quantum communication, *Nature Photonics* **1,** 165 - 171 (2007)

5. Salter, C. L., Stevenson, R. M., Farrer, I., Nicoll, C. A., Ritchie, D. A.& Shields, A. J. An entangled-light-emitting diode, *Nature* **465,** 594–597 (2010)

6. Gong, M., et al. Statistical properties of exciton fine structure splitting and polarization angles in quantum dot ensembles, *Phys. Rev. B* **89**, 205312 (2014)

7. Aspect, A., Grangier, P. & Roger, G. Experimental Tests of Realistic Local Theories via Bell's Theorem, *Phys. Rev. Lett.* **47**, 460 (1981)

8. Benson, O., Santori, C., Pelton, M. & Yamamoto, Y. Regulated and Entangled Photons from a Single Quantum Dot, *Phys. Rev. Lett.* **84,** 2513 (2000)

9. Gammon, D., et al. Fine Structure Splitting in the Optical Spectra of Single GaAs Quantum Dots, *Phys. Rev. Lett.* **76,** 3005 (1996)

10. Bayer, M. et al. Fine structure of neutral and charged excitons in self-assembled In(Ga)As/(Al)GaAs quantum dots, *Phys. Rev. B* **65,** 195315 (2002)

11. Bennett, A. J. et al. Electric-field-induced coherent coupling of the exciton states in a single quantum dot, *Nat. Physics* **6**, 947 – 950 (2010)

12. R. Trotta, et al. Universal Recovery of the Energy-Level Degeneracy of Bright Excitons in InGaAs Quantum Dots without a Structure Symmetry, *Phys. Rev. Lett.* **109,** 147401 (2012)

13. R. Trotta, et al. Highly Entangled Photons from Hybrid Piezoelectric- Semiconductor Quantum Dot Devices, *Nano Lett.* **14**, 3439 – 3444 (2014)

14. Rastelli, A. et al. *In situ* laser microprocessing of single self-assembled quantum dots and optical microcavities, *Appl. Phys. Lett.* **90,** 073120 (2007)

15. Faraon, A., Englund, D., Fushman, I., Vučković, J., Stoltz, N., Petroff & P. Local quantum dot tuning on photonic crystal chips, *Appl. Phys. Lett.* **90,** 213110 (2007)

16. Young, R. J. et al. Inversion of exciton level splitting in quantum dots, *Phys. Rev. B* **72,** 113305 (2005)

17. Trotta, R., Martín-Sánchez, J., Daruka, I., Ortix, C., & Rastelli, A. Energy-Tunable Sources of Entangled Photons: A Viable Concept for Solid-State-Based Quantum Relays, *Phys. Rev. Lett.* **114,** 150502 (2015)

18. Gong, M., Zhang, W., Guo, G.-C. & He, L. X. Exciton Polarization, Fine-Structure Splitting, and the Asymmetry of Quantum Dots under Uniaxial Stress, *Phys. Rev. Lett.* **106,** 227401 (2011)

19. Zhang, J. X. et al. A Nanomembrane-Based Wavelength-Tunable High-Speed Single-Photon-Emitting Diode, *Nano Lett.* **13,** 5808–5813 (2013)

20. Trotta, R. et al. Nanomembrane Quantum-Light-Emitting Diodes Integrated onto Piezoelectric Actuators, *Adv. Mat.* **24,** 2668 – 2672 (2012)

21. Han, P., Yan, W., Tian, J., Huang, X. & Pan, H. Cut directions for the optimization of piezoelectric coefficients of lead magnesium niobate–lead titanate ferroelectric crystals, *Appl. Phys. Lett.* **86,** 052902 (2005)

22. Plumhof, J. D. et al. Strain-induced anticrossing of bright exciton levels in single self-assembled GaAs/Al$_x$Ga$_{1-x}$As and In$_x$Ga$_{1-x}$As/GaAs quantum dots, *Phys. Rev. B* **83,** 121302(R) (2011)

23. Akopian, N. et al. Entangled Photon Pairs from Semiconductor Quantum Dots, *Phys. Rev. Lett.* **96,** 130501 (2006)

24. Avron, J. E., Bisker, G., Gershoni, D., Lindner, N. H., Meirom, E. A. & Warburton, R. J. Entanglement on Demand through Time Reordering, *Phys. Rev. Lett.* **100,** 120501 (2008)

25. James, D. F. V., Kwiat, P. G., Munro, W. J. & White, Andrew G. Measurement of qubits,

*Phys. Rev. A* **64,** 052312 (2001)

26. Michler, P. Single Semiconductor Quantum Dots, Springer, 2009.

27. Hudson, A. J. et al. Coherence of an Entangled Exciton-Photon State, *Phys. Rev. Lett.* **99,** 266802 (2007)

28. Stevenson, R. M. et al. A semiconductor source of triggered entangled photon pairs, *Nature* **439,** 179-182 (2006)

29. Dousse, A. et al. Ultrabright source of entangled photon pairs, *Nature* **466,** 217–220 (2010)

30. Juska, G. et al., Towards quantum-dot arrays of entangled photon emitters, *Nature Photonics* **7**, 527–531 (2013)

31. Young, R. J., et al. Bell-Inequality Violation with a Triggered Photon-Pair Source, *Phys. Rev. Lett.* **102,** 030406 (2009)

32. Marijn A. M. et al. Observation of strongly entangled photon pairs from a nanowire quantum dot, *Nature Communications*, **5**, 5298 (2014)

**Acknowledgement**

The work was supported financially by BMBF QuaHL-Rep (Contracts no. 01BQ1032 and 01BQ1034), Q.Com-H (16KIS0106) and the European Union Seventh Framework Programme 209 (FP7/2007-2013) under Grant Agreement No. 601126 210 (HANAS). J. X. Zhang was supported by China Scholarship Council (CSC, No. 2010601008). The authors thank B. Eichler, R. Engelhard, P. Atkinson, B. Höfer and S. Harazim for discussions and the technical support.

## Author contributions

The sample was grown by Y. H. and E. Z. and the device was fabricated by J. X. Z. The work was conceived and designed by J. X. Z., F. D. and R. T., and guided by A. R. and O. G. S. The optical measurements were performed by J. X. Z., J. W., R. T. and D. H. The results were discussed by all authors. J. X .Z wrote the manuscript with help from all the other authors.

## Additional information

The authors declare no competing financial interests. Supplementary information accompanies this paper on [www.nature.com/naturephotonics](www.nature.com/naturephotonics). Reprinting and permissions information is available online at [http://npg.nature.com/reprintsandpermissions](http://npg.nature.com/reprintsandpermissions). Correspondence and requests for materials should be addressed to J. X. Z., F. D and R. T.

# Supplementary Information

# Strain-tunable entangled-light-emitting diodes with high yield and fast operation speed

*Jiaxiang Zhang [1] [\*], Johannes S. Wildmann [2], Fei Ding [1] [\*], Rinaldo Trotta [2] [\*], Yongheng Huo [1], Eugenio Zallo [1], Daniel Huber [2], Armando Rastelli [2] & Oliver G. Schmidt [1]*

[1] Institute for Integrative Nanosciences, IFW Dresden, Helmholtzstraße 20, 01069 Germany

[2] Institute of Semiconductor and Solid State Physics, Johannes Kepler University Linz, Altenbergerstraße 69, 4040, Austria

## S1: Fine structure splitting (FSS, *s*) and exciton polarization direction $\theta$: theoretical background

The effect of our anisotropic stress applied to the QDs can be seen as two stresses $p_{xx}$ and $p_{yy}$ along the [1-10] and [110] crystal directions, respectively. Note that the behaviors of FSS are almost symmetric for stresses along the [110] and [1-10] directions; i.e., the effects of tensile stress along the [110] direction are almost identical to the effects of compression along the [1-10] direction [1]. Therefore, our anisotropic stress can be reliably treated as a 'uniaxial stress' along one of the principle stress axes ([110] or [1-10] direction). The behavior of *s* and $\theta$ can be described using the following formulas [1]:

$$s = \sqrt{4(\beta p + k)^2 + (\alpha p + 2\delta)^2} \quad (1)$$

$$\theta = tan^{-1}\left(\frac{-2\delta - p\alpha \pm s}{2(\beta p + k)}\right)$$

(2)

Where *p* is the magnitude of the external stress and it is determined by $p = p_{yy} - p_{xx}$; $\beta$ and $\alpha$ are stress dependent parameters, and $k$, $\delta$ depends on the microscopic structure of the QDs. For stress along either [110] or [1-10] direction, $\beta \approx 0$ and the minimum FSS is determined by $s_{min} = 2|k| = s_0 \times \sin(2\theta_0)$ ($s_0$ and $\theta_0$ are the FSS and polarization angle at zero stress, see Fig. S1). In addition, $k$, $\delta$ can be experimentally determined by $k = -s_0 \times \sin(2\theta_0)/2$, $\delta = s_0 \times \cos(2\theta_0)/2$. Table 1 is a summary of $s_0$, $\theta_0$, $k$, $\delta$ for the five studied dots in the main text. With these parameters, we can fit the experimental data theoretically using formulas (1) and (2) above, as the solid lines shown in Fig. 2 in the main text.

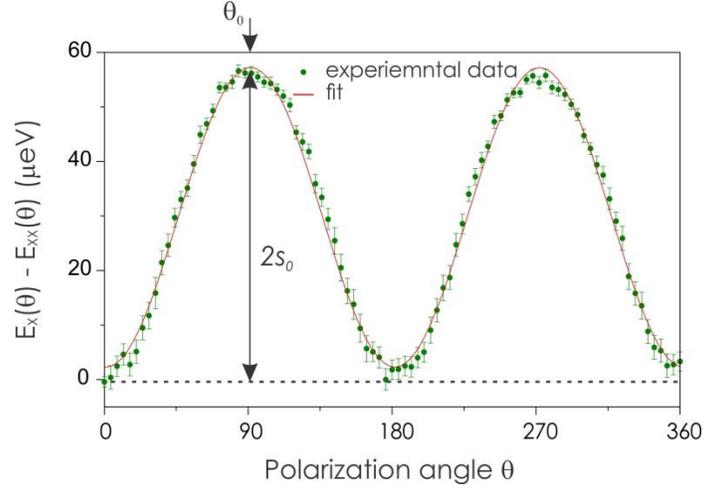

**Figure S1:** The result of a typical measurement of the $E_X(\theta)$ - $E_{XX}(\theta)$ from which we can extract $s_0$ and $\theta_0$. $\theta$ used here, as well as in the main text, is the polarization angle of high-energy component of the exciton emission.

**Table S1: summary of $s_0, \theta_0, k, \delta$ for the studied five dots**

| QDs | $\theta_0$ (°) | $s_0$ (μeV) | κ (μeV) | δ (μeV) |
|---|---|---|---|---|
| A | 102.0±0.4 | 20.1±0.2 | 4.11±0.1 | -9.20±0.08 |
| B | 92.7±0.2 | 34.9±0.2 | 1.60±0.08 | -17.4±0.1 |
| C | 86.1±0.4 | 17.31±0.2 | -1.20±0.05 | -8.5±0.1 |
| D | 90.4±0.3 | 16.9±0.1 | 0.12±0.05 | -8.40±0.07 |
| E | 90.6±0.3 | 27.4±0.3 | 0.28±0.07 | -13.7±0.1 |

**S2: Statistics of the initial polarization angle $\theta_0$ and FSS $s_0$**

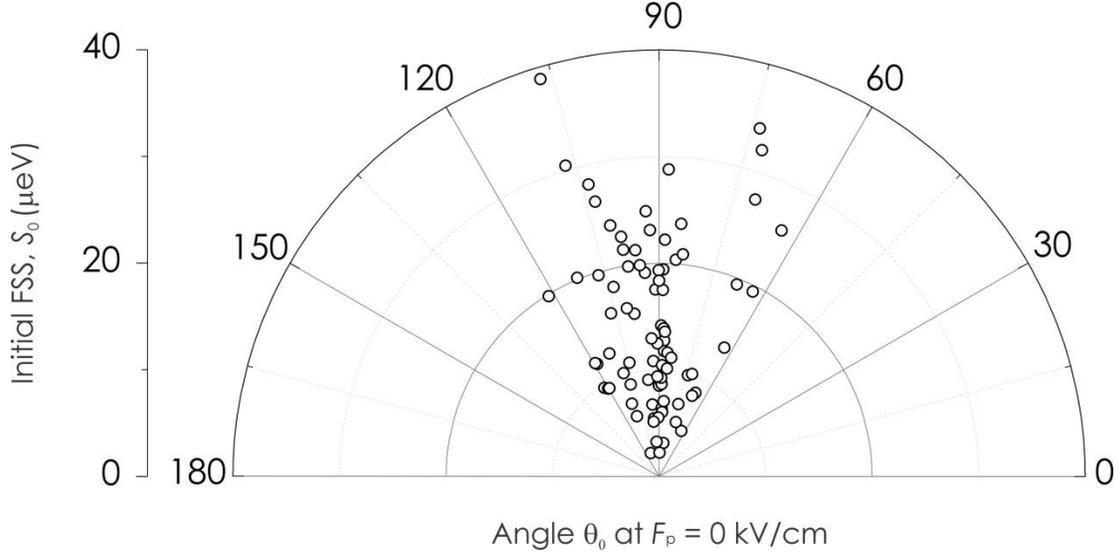

**Figure S2:** Scatter plot of the polarization angle $\theta_0$ at zero strain fields relative to the [110] direction of the GaAs [2]. The data is obtained by characterizing 82 randomly selected QDs in our S-ELED device and each dot has distinct exciton and biexciton emission. It is clearly seen that the majority of dots have $\theta_0$ at about $90^o$, that is, the [1-10] crystal axis of GaAs. Some scatter of $\theta_0$ in a range of $60^o \sim 120^o$ has been also observed which is likely due to the anisotropy in strain and composition during the growth process. Meanwhile, a large number of dots have small deviations from the [1-10] crystal axis, which can provide a tolerance to the alignment of the strain direction with respect to the crystal axes. Our statistical investigation presented here is in line with the results reported in ref. 2. In addition, the initial FSS $s_0$ was revealed to have average value of about 20 μeV, and the dots with sufficient small $s_0$ for entanglement are barely found without applying any strain fields.

**S3: Entanglement analysis for other dots at $s_{\min} < 1$ μeV**

In addition to the QD E characterized in the main text, here we present an entanglement characterization for additional three dots. Table S2 summarizes the entanglement results for these other dots in terms of the fidelity to the maximally entangled Bell state $|\Psi^+\rangle = 1/\sqrt{2}\,(|H_{XX}H_X\rangle + |V_{XX}V_X\rangle)$. The dot 1 is the dot D studied in the main text and its strain tuning behavior is shown in Fig. 2a and 2b (red curves). The change of the FSS with the external strain fields for the other two dots is shown in Fig. S3. By measuring the degree of correlation in given basis at the minimum FSS, the fidelity of these dots is found to be $0.72\pm0.03$, $0.75\pm0.02$ and $0.71\pm0.02$ respectively. Such high values of fidelity exceed the classical limit of 0.5 and suggest all these dots have been successfully tuned to be 'good' entangled light sources via the external strain field.

**Table S2 | Summary of entanglement results for other three dots.**

| dots | X (eV) | XX (eV) | $V_d$ (V) | $s_{min}$ (µeV) | $f^+$ |
|---|---|---|---|---|---|
| 1 | 1.4127 | 1.4122 | -1.75 | 0.30 ±0.25 | 0.72±0.03 |
| 2 | 1.4013 | 1.4000 | -1.98 | 0.20±0.30 | 0.75±0.02 |
| 3 | 1.4000 | 1.3993 | - 2.05 | 0.60±0.20 | 0.71± 0.02 |

The binding energy of exciton and biexciton were measured at $F_p = 0$ kV·cm$^{-1}$ and all the dots are characterized under d. c. current injection.

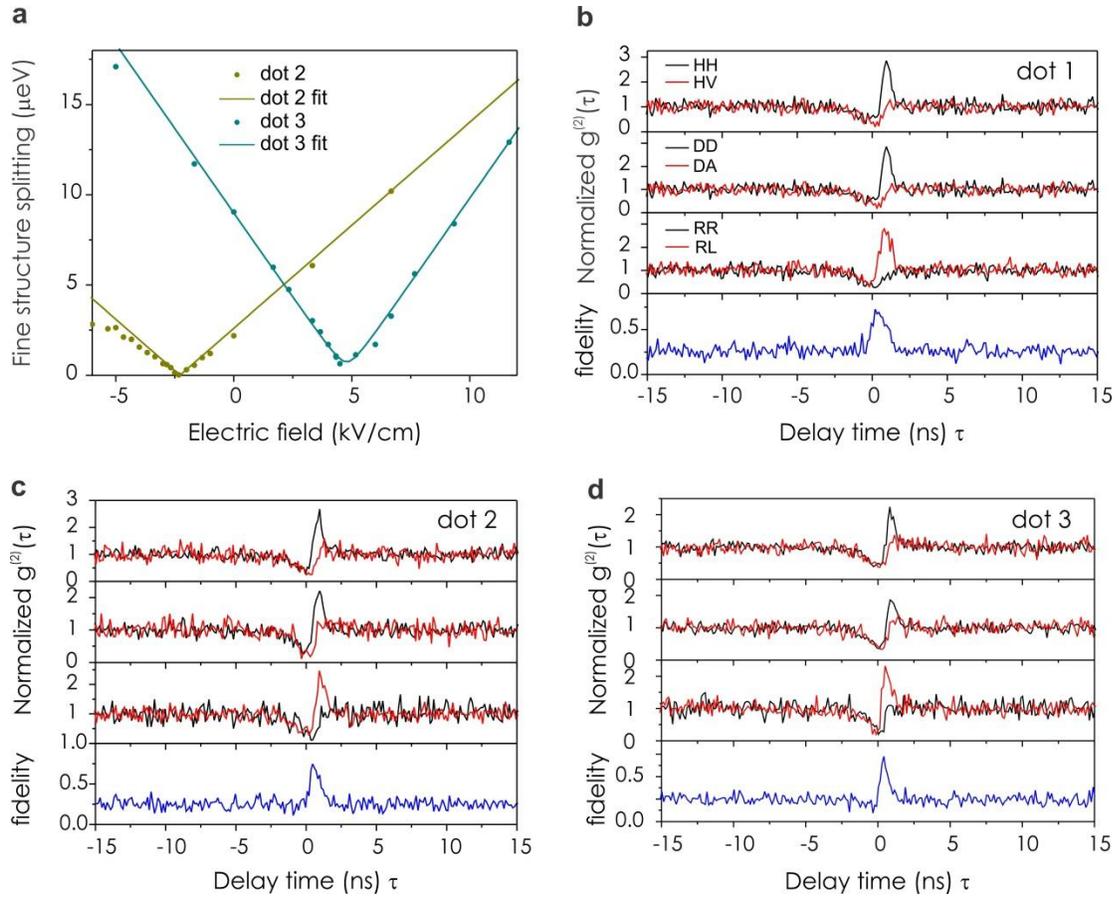

**Figure S3:** **(a)** FSS as function of the electric field $F_p$ applied on the PMN-PT for the dot 2 and dot 3, respectively. **(b) – (d)** The polarization correlation results measured in rectilinear (HV), diagonal (DA) and circular (RL) bases for the dot 1, 2, 3. It is clearly seen that, for co-polarized exciton and biexciton photons (black curves), correlation is always observed in HV(DA) basis and anti-correlation in RL basis. The fidelity is found to be $0.71\pm0.03$, $0.75\pm0.02$ and $0.71\pm0.02$ for the dot 1, 2 and 3, respectively.

## S4: Time gating and violation of Bell's inequality

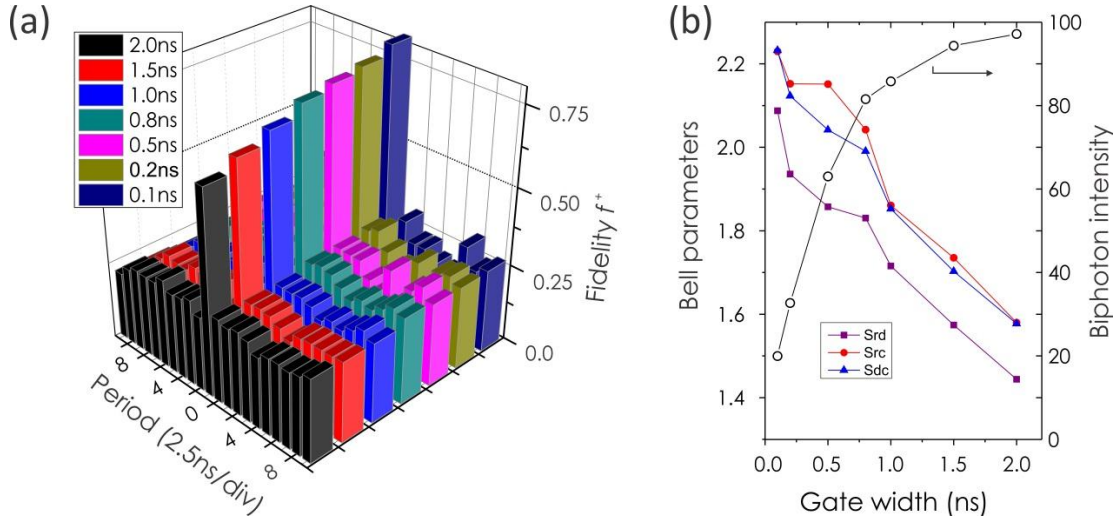

**Figure S4**: **(a)** Fidelity as a function of the temporal gate width, which is extracted from the co- and cross-polarization correlations at the excitation repetition rate of 400 MHz as shown in Fig. 5a in the main text. By narrowing the gate width from 2.0 to 0.1 ns, the fidelity increases drastically from $0.67\pm0.02$ to $0.83\pm0.05$. **(b)** Gate width dependent change of Bell parameters ($S_{RD}$, $S_{RC}$ and $S_{DC}$) and the biphoton intensity. In our experiment, we can clearly see that all the three Bell parameters increase rapidly as the gate width decreases. $S_{RD}$ is found smaller than $S_{RC}$ and $S_{DC}$ due to the effect of the fluctuating magnetic fields produced by the QD nuclei, as pointed out by Stevenson et al. [3]. Therefore, for actual entangled state, $S_{RD}$ is not necessarily optimized to test Bell's inequality [4, 5]. We found that $S_{RC}$ and $S_{DC}$ become larger than 2 as the gate width is narrowed down to 0.8 ns, while for a gate width of 0.1 ns all the three Bell's parameters shown violation of Bell's inequality by at least one standard deviation. Furthermore, the proportion of the total biphoton intensity as a function of the time gate width is also shown. Although the fidelity and the Bell parameter increase as we decrease the gate width, we also note a drastic decrease of the biphoton intensity as the gate width is decreased. At the gate width of 2.5 ns, all the coincidence counts are included for calculating the degree of correlation (see in the main text) and thus it corresponds to a biphoton intensity proportion of 100%. As smaller gate width is applied from 2.0 ns to 0.1 ns, the biphoton intensity is reduced quickly from 97% to 16%.

**References**

1. Gong, M., Zhang, W., Guo, G.-C. & He, L. X. Exciton Polarization, Fine-Structure Splitting, and the Asymmetry of Quantum Dots under Uniaxial Stress, *Phys. Rev. Lett.* **106,** 227401 (2011)

2. Bennett, A. J. et al. Electric-field-induced coherent coupling of the exciton states in a single quantum dot, *Nat. Physics* **6**, 947 – 950 (2010)

3. R. M. Stevenson. et al., http://arxiv.org/abs/1103.2969

4. Salter, C. L., Stevenson, R. M., Farrer, I., Nicoll, C. A., Ritchie, D. A. & Shields, A. J. An entangled-light-emitting diode, *Nature* **465,** 594–597 (2010)

5. R. Trotta, et al. Highly Entangled Photons from Hybrid Piezoelectric- Semiconductor Quantum Dot Devices, *Nano Lett.* **14**, 3439 – 3444 (2014)